\documentclass[12pt,preprint]{aastex}

\shorttitle{XMM-Newton observations of SGR 1900$+$14}

 \shortauthors{Mereghetti  et al.}

\def \src {SGR\thinspace1900$+$14}
\def\approxgt{\mathrel{\hbox{\rlap{\lower.55ex \hbox {$\sim$}}
        \kern-.3em \raise.4ex \hbox{$>$}}}}
\def\approxlt{\mathrel{\hbox{\rlap{\lower.55ex \hbox {$\sim$}}
        \kern-.3em \raise.4ex \hbox{$<$}}}}

\def\flux {\mbox{erg cm$^{-2}$ s$^{-1}$}}
\def\lum {\mbox{erg s$^{-1}$}}

\begin{document}


\title{The first XMM-Newton observations of the Soft Gamma-ray 
Repeater SGR 1900$+$14 \footnote{Based on obervations obtained with XMM-Newton,
an ESA science mission with instruments and contributions
directly funded by ESA Member States and NASA.}}


\author{S. Mereghetti, P. Esposito\altaffilmark{2}, A. Tiengo}
\affil{INAF - IASF Milano, via Bassini 15, I-20133 Milano, Italy}

\author{S. Zane}
\affil{Mullard Space Science Laboratory, University College
London, \\ Holmbury St. Mary, Dorking Surrey, RH5 6NT, United Kingdom}

\author{R. Turolla}
\affil{Universit\`a di Padova, Dipartimento di Fisica, via Marzolo
8, I-35131 Padova, Italy}

\author{L. Stella, G.L. Israel}
\affil{INAF - Osservatorio Astronomico di Roma,
 via Frascati 33, \\ I-00040 Monteporzio Catone, Italy}

\author{ D. G\"{o}tz}
\affil{CEA Saclay, DSM/DAPNIA/Service d'astrophysique, F-91191 Gif-sur-Yvette, France}

\author{M. Feroci}
\affil{INAF - IASF Roma, v. Fosso del Cavaliere 100, I-00133 Roma,
Italy }


\altaffiltext{2}{Universit\`a di Pavia, Dipartimento di Fisica Nucleare e Teorica and INFN-Pavia, via Bassi 6, I-27100 Pavia, Italy}


\begin{abstract}
A $\sim$50 ks  XMM-Newton observation of  \src\ has been carried
out in September 2005, after almost three years during which no
bursts were detected from this soft gamma-ray repeater. The 0.8-10
keV spectrum was well fit by a power law plus blackbody model with
photon index $\Gamma$=1.9$\pm$0.1, temperature kT=0.47$\pm$0.02
keV and N$_H$ = $(2.12\pm0.08)\times10^{22}$ cm$^{-2}$, similar to
previous observations of this source. The flux was
$\sim5\times10^{-12}$ \flux , a factor 2 dimmer than the typical
value and the smallest ever seen from \src . The long term fading
of the persistent emission has been interrupted by the recent
burst reactivation of the source. A target of opportunity
XMM-Newton observation performed in April 2006 showed a flux
$\sim$15\% higher. This variation was not accompanied by
significant changes in the spectrum, pulsed fraction and light
curve profile. We searched for emission and absorption lines in
the spectra of the two observations, with negative results and
setting tight upper limits of 50--200 eV (3$\sigma$), depending on
the assumed line energy and width, on the equivalent width of
lines in the 1-9 keV range.
\end{abstract}





\section{Introduction}

The few  X-/gamma-ray sources known as Soft Gamma-ray Repeaters
(SGRs) are generally believed to provide the most convincing
evidence for the existence of magnetars, i.e. neutron stars with
magnetic fields well above the quantum critical value  $B_{c} =
\frac{m_{e}^{2}c^{3}}{e\hbar}=4.4\times10^{13}$~G \citep{duncan92,
paczynski92}.

The first sources of this class were discovered in the seventies
as transient emitters of short (tens of ms) bursts of hard X-rays
\citep{mazets79,cline80,laros86}. Only in more recent years it was
possible to identify and study in detail their counterparts in the
classical 1--10~keV X-ray band. This led to the discovery of
periodic pulsations and secular spin-down in SGR 1806--20 (P=7.5
s, \citealt{kouveliotou98}) and \src\ (P=5.2 s,
\citealt{hurley99,kouveliotou99}) thus confirming the neutron star
nature of this class of sources. Occasionally, SGRs emit very
energetic giant flares, during which up to a few 10$^{46}$ ergs
are released in a few tenths of a second. The extreme properties
of these events, of which only three, each one for a different
source, have been observed to date, are the main motivation for
the magnetar interpretation \citep{thompson95,thompson96}.

Among the four confirmed SGRs, SGR 1806--20 and \src\, offer the
best prospects for detailed spectral and timing studies in the
soft X-ray band (E$<$10 keV). Both these sources, despite
alternating periods of bursting activity with intervals of
quiescence lasting months or years, remained at flux levels of the
order of 10$^{-11}$ \flux. The other Galactic soft repeater, SGR
1627--41, had a similar X-ray flux when it was discovered as a
bursting source in 1998 \citep{woods99c}, but since then its
luminosity decreased by a factor $\sim$25 and it is now a rather
faint source \citep{mereghetti06}. SGR 0526--66 has a relatively
high luminosity ($\sim2\times10^{35}$ \lum), but being at the
Large Magellanic Cloud distance, its  observed flux is only
$\sim$10$^{-12}$ \flux\ \citep{kulkarni03}. Furthermore, its study
with low spatial resolution instruments is complicated by the
presence of diffuse emission from the surrounding supernova
remnant N49.

We started in 2003 a long--term monitoring program to study the
time evolution of the spectral properties of SGR 1806--20 using
the XMM-Newton X-ray satellite. Thanks to its imaging capability
and large effective area we could obtain spectra of a much better
quality than those available from previous satellites. The
emission of the 27 December 2004 giant flare was also a fortunate
occurrence, since we could observe how the source properties
evolved in the two years leading to the flare and how they changed
after this dramatic event \citep{mte05,tiengo05}.

A similar observational campaign could not be carried out for
\src. In fact this source is located in a sky region that, until
recently, was not accessible to XMM-Newton observations due to
technical constraints in the satellite pointing. Thus the first
observation of \src\ with this satellite could   be obtained only
in September 2005, during a long period  of inactivity (the last
bursts before the XMM-Newton observation were reported in November
2002, \citealt{hurley02}).  \src\ became active again in March
2006: bursts were observed by Swift \citep{palmer06} and
Konus-Wind \citep{golenetskii06}. We therefore requested a target
of opportunity XMM-Newton observation, that was carried out on 1
April 2006.

\section{Observations}

We present the results  obtained with the EPIC instrument,
consisting of two MOS and one pn cameras
\citep{turner01,struder01}. In all the observations the pn was
operated in Full Frame mode and the MOSs in Large Window mode
(time resolution: 73.4 ms and 0.9 s respectively). Both the pn and
the MOSs mounted the medium thickness filter. All the data were
processed using version 6.5.0 of the \emph{XMM-Newton Science
Analysis System} and the most recent calibration files (last
update on 2005 December 14). Response matrices and effective area
files were generated ad-hoc with the  SAS tasks \emph{rmfgen} and
\emph{arfgen}; spectral fits were performed using the XSPEC
v11.3 software \citep{arnaud96}.

The first  observation of \src\ was divided in two parts, starting
on 2005 September 20 01:44 UT  and 22 01:36 UT, respectively.
Since there was no evidence for variations in the flux and
spectrum of the source from September 20 to 22, we added the two
data sets and analyzed them together. After filtering for particle
induced flares we obtained a net exposure time of 38.9 ks in the
pn camera, and of 47.4 ks in the two MOSs.

The second observation started on 1 April 2006 and lasted $\sim$22
ks, yielding net exposure times of 12.7 ks in the pn camera and of
15.7 ks in the two MOSs.

\begin{figure}
\includegraphics[angle=00,width=15cm]{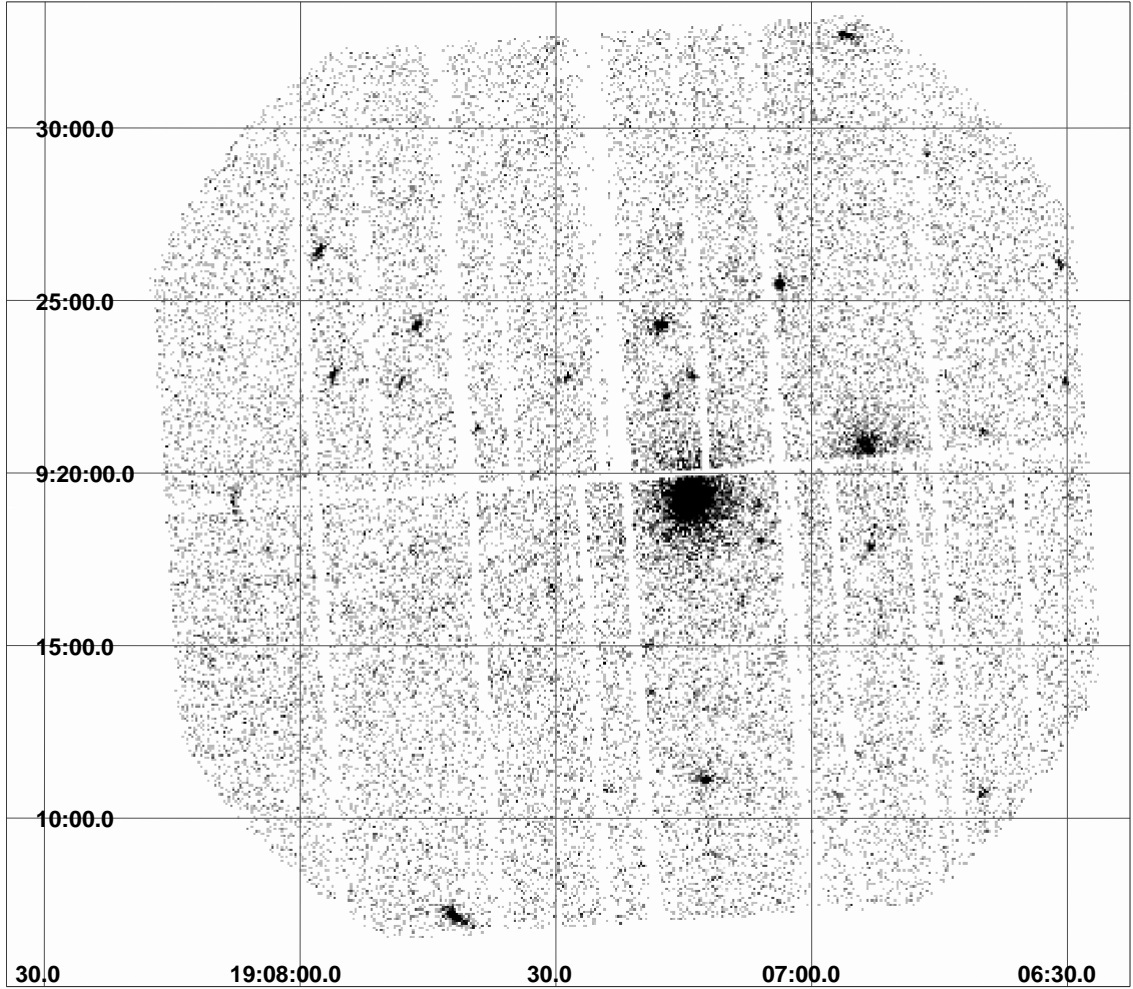}
\caption{\label{ds9} EPIC-pn image of the \src\ field in the
0.8--10 keV energy range. North is to the top, East to the left.}
\end{figure}

The 0.8--10 keV image obtained with the pn camera in September
2005 is shown in Fig.\ref{ds9}. \src\
is the brightest source at the
center of the field. Several other objects, detected here for the
first time are visible. As expected for such a low Galactic
latitude field, many of them can be associated with foreground
stars based on their soft spectrum and positional coincidence with
bright optical counterparts. A relatively bright spatially
resolved source is also visible $\sim5'$ to the West of \src , but
it is very likely unrelated to the SGR. Its spectrum is well
described by an optically thin plasma emission model (MEKAL in
XSPEC) with temperature \mbox{$kT=7^{+3}_{-2}$ keV} and a high
absorption of $N_{\rm H}=(3.6^{+1.0}_{-0.7})\times10^{22}$
cm$^{-2}$. This spectrum and the spatial extension of about one
arcminute are consistent with emission from a cluster of galaxies
at redshift $z\sim0.6$ and with a 2--10 keV luminosity of
$\sim$$2\times10^{44}$ \lum. Its coordinates are R.A.=19$^{h}$
06$^{m}$ 53$^{s}$.7, Dec.=+09$^{\circ}$ 20$'$ 47$''$ (J2000).

\section{Timing and spectral results}

\begin{figure}
\includegraphics[angle=90,width=15cm]{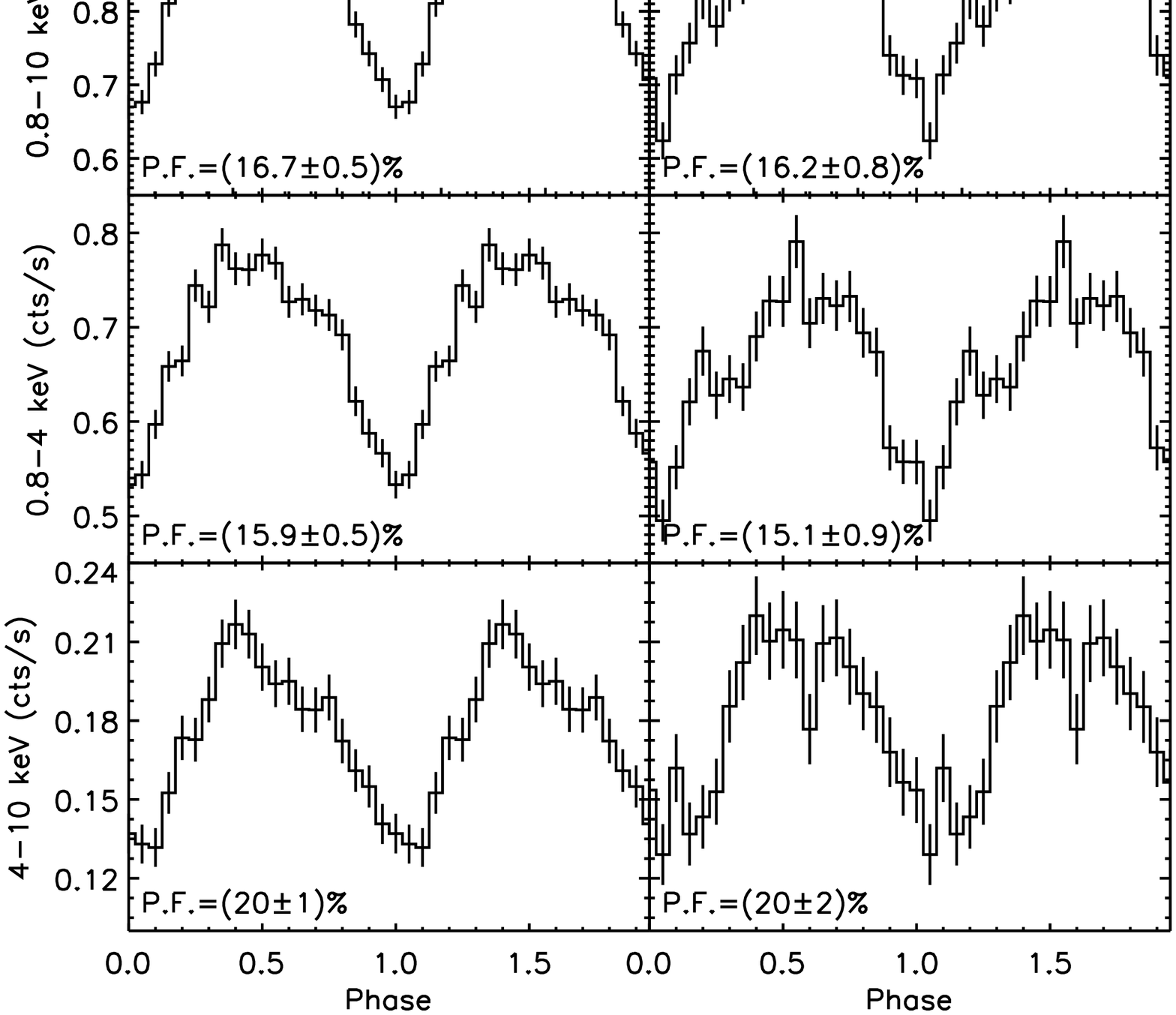}
\caption{\label{fol}Folded light curves in the total (0.8-10 keV),
soft (0.8-4 keV), and hard (4-10 keV) energy range for the two
observations. The background has been subtracted. The
corresponding pulsed fraction is indicated on each panel
(1$\sigma$ errors).
 }
\end{figure}

Except for the periodic pulsations, \src\ did not show flux
variability within the two observations, but it was about 15\%
brighter in April 2006, after the burst reactivation. We searched
for the presence of bursts in both observations, by a careful
analysis of light curves binned with different time resolution,
but none could be found. With a standard folding analysis of the
Solar system barycentered  light curves, we measured a spin period
of \mbox{$5.198346\pm0.000003$ s} in September 2005 and
\mbox{$5.19987\pm0.00007$ s} in April 2006. In Figure \ref{fol} we
show the background subtracted pulse profiles in three different
energy ranges. The pulsed fractions (values reported in the
corresponding figures) have been computed by fitting a sinusoid to
the light curves.  There is no evidence for changes in the pulsed
fractions and light curve shapes between the two observations. The
two period measurements correspond to a spin-down rate of
(9.2$\pm$0.4)$\times10^{-11}$ s s$^{-1}$.

We extracted spectra for \src\ by selecting source counts  with
patterns 0--4 for the pn camera and 0--12 for the MOS cameras from
circles of 40\arcsec\ radius. The background spectra were
extracted from composite regions located on the same chip as the
source. The spectra were rebinned to have at least 30 counts in
each bin and to oversample the instrumental energy resolution by a
factor three. Fits were performed in the energy range 0.8--12 keV,
since the source is heavily absorbed and only few counts are
detected at lower energies.

\begin{figure}
\includegraphics[angle=270,width=15cm]{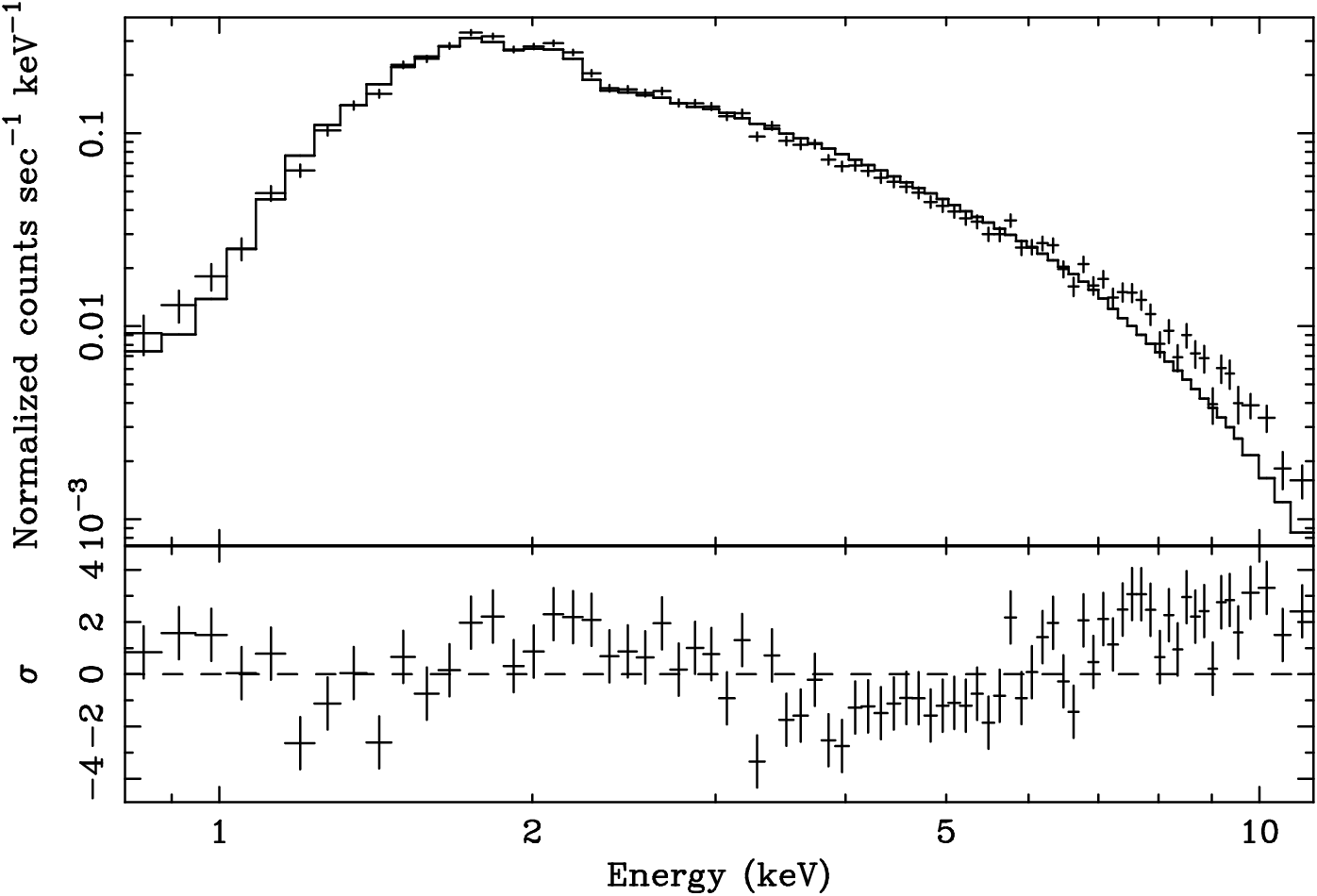}
\caption{\label{spec1} EPIC pn spectrum of \src\ from the
September 2005 observation. Top: data and best fit power law
model. Bottom: residuals from the best fit model in units of
standard deviations.}
\end{figure}

\begin{figure}
\includegraphics[angle=270,width=15cm]{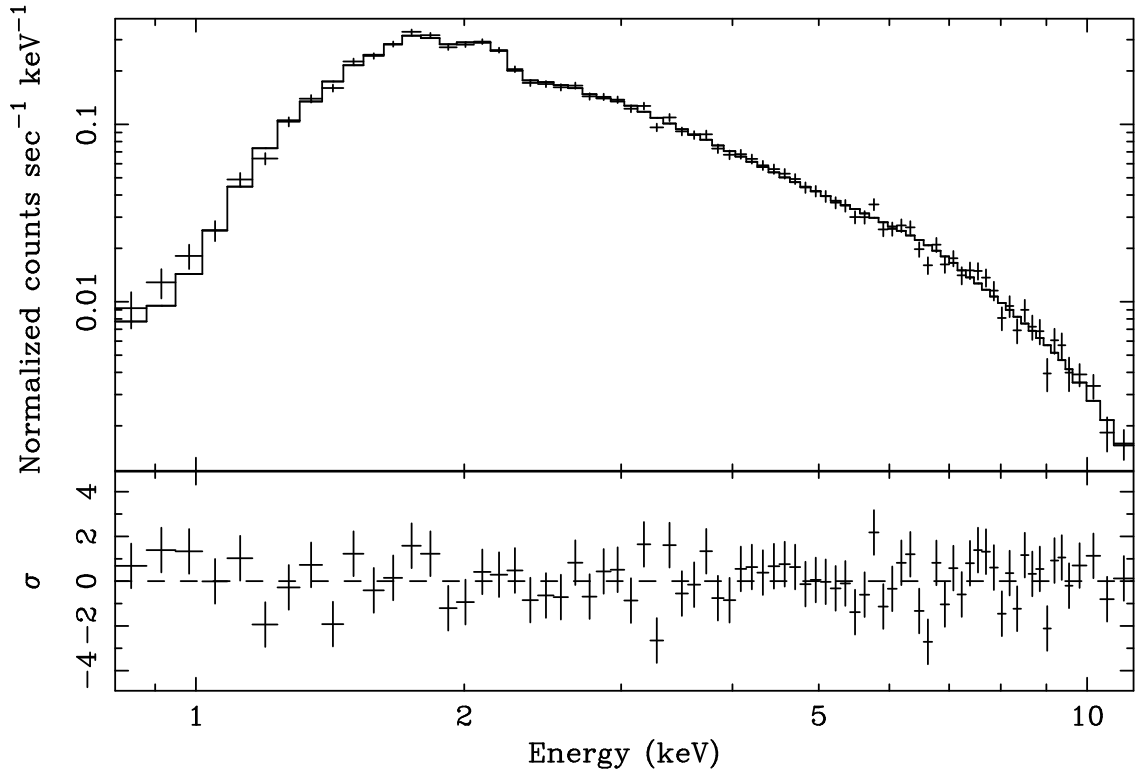}
\caption{\label{spec2} EPIC pn spectrum of \src\ from the
September 2005 observation. Top: data and best fit power law plus
blackbody model. Bottom: residuals from the best fit model in
units of standard deviations.}
\end{figure}

In Fig.~\ref{spec1} and \ref{spec2} we show the spectrum obtained
with the pn camera in the September 2005 observation, fitted with
a power law and with a power law plus black body model,
respectively. The latter clearly provides a better fit, as it can
be seen from the residuals shown in the lower panels of the
figures. Similar results were obtained using the spectra from the
MOS. We therefore performed simultaneous fits of the spectra from
the three cameras, obtaining photon index $\Gamma$=1.9$\pm$0.1,
blackbody temperature kT=0.47$\pm$0.02 keV, and absorption N$_H$ =
$(2.12\pm0.08)\times10^{22}$ cm$^{-2}$. An acceptable fit could
also be obtained with the sum of two blackbodies with temperatures
of 0.53 and 1.9 keV (see Table~\ref{fits}).

The second observation gave entirely consistent spectral
parameters, except for a statistically significant variation in
the flux. The background subtracted count rates (0.8-10 keV) measured with the
pn camera were  $0.615\pm0.004$ counts s$^{-1}$ in September 2005
and  $0.720\pm0.008$ counts s$^{-1}$ in April 2006. 
Indeed the April 2006 data are well described simply
rescaling in normalization (by $\sim$15\%) the best fit spectra of
the September 2005 observation.

For both observations we performed  phase-resolved spectroscopy
extracting the  spectra for different selections of phase
intervals. No significant variations with phase were detected, all
the spectra being consistent with the model and parameters of the
phase-averaged spectrum, simply rescaled  in normalization.

\begin{figure}
\includegraphics[angle=90,width=13cm]{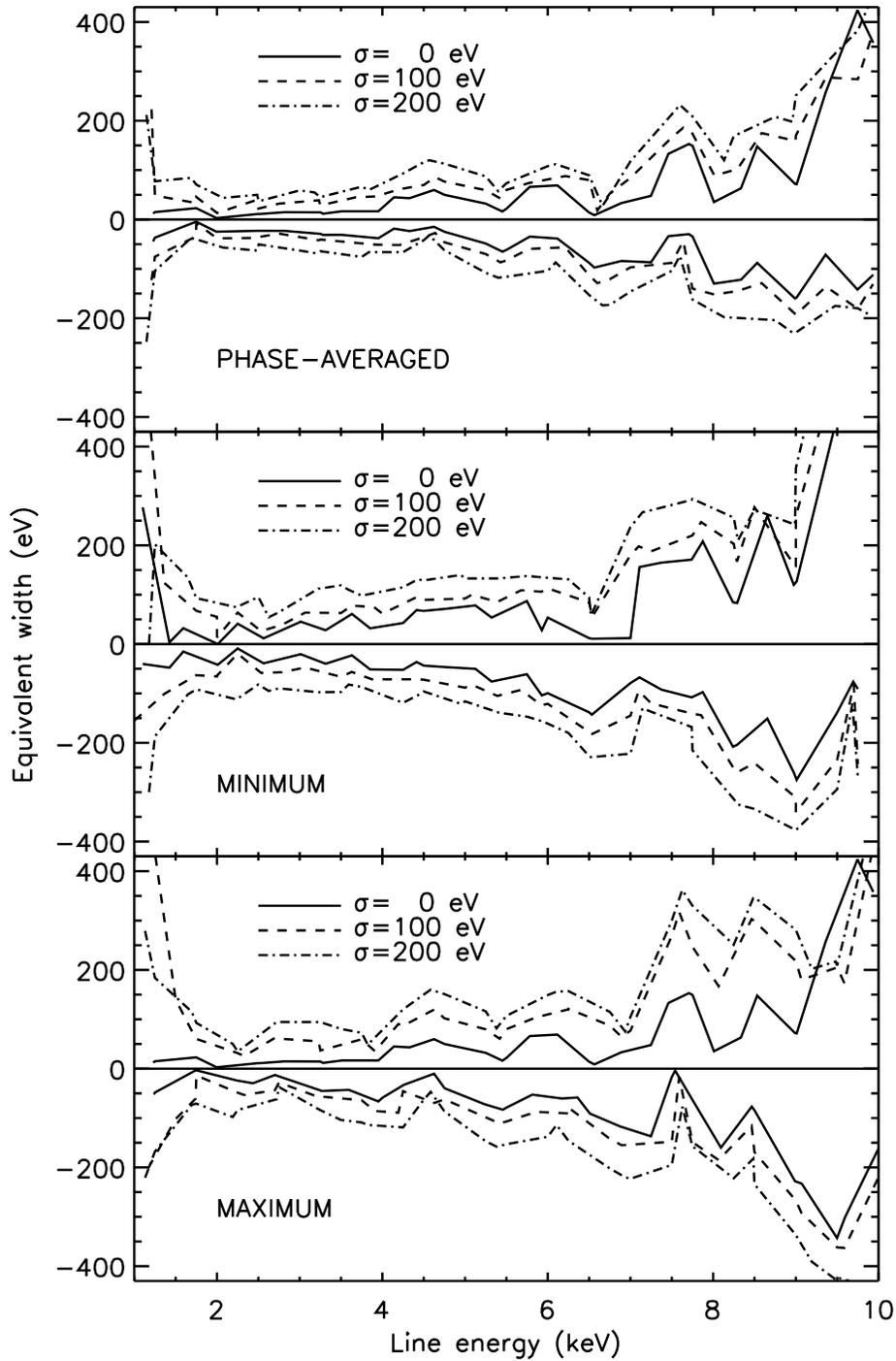}
\caption{\label{lines} Upper limits (at 3 $\sigma$) on spectral
features in the 2005 pn data of \src\ . The top panel refers to
the phase-averaged spectrum and the two lower panels to the
spectra at the pulse minimum and  maximum.
 }
\end{figure}

No evidence for emission or absorption lines was found by
inspecting the residuals from the best fit models. We computed
upper limits on the lines equivalent widths as a function of the
assumed line energy and width. This was done by adding Gaussian
components to the model and computing the allowed range in their
normalization. The most constraining results were obtained for the
September 2005 observation. They are summarized in Figure
\ref{lines}, where the top panel refers to the phase averaged
spectrum and the other ones to the spectra of the pulse maximum
(phase from 0.25 to 0.75 of Fig.~\ref{fol}) and minimum.

\section{Discussion}

\begin{figure}
\includegraphics[angle=0,width=15cm]{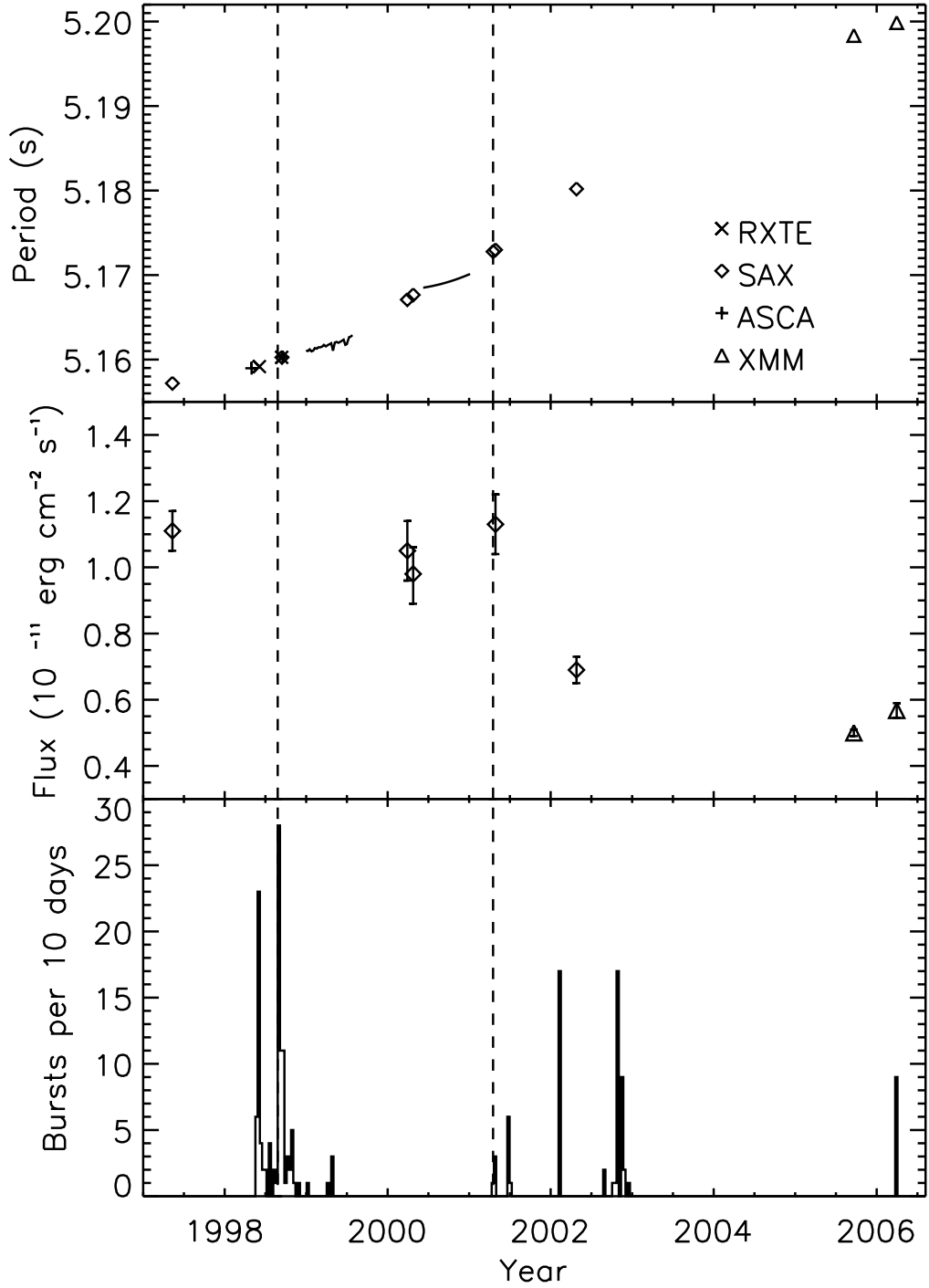}
\caption{\label{multi} Long term evolution of the pulse period
(top panel), X--ray flux (middle panel) and bursting rate observed with 
the IPN (bottom
panel) of SGR 1900+14. The vertical dashed lines indicate the
times of the 27 August 1998 giant flare and of the 18 April 2001
intermediate flare. }
\end{figure}

Previous spectral studies of the persistent X-ray emission of
\src, carried out with ASCA \citep{hurley99}, BeppoSAX (e.g.,
\citealt{woods99b}, \citealt{esposito06} and references therein)
and Chandra \citep{kouveliotou01}, showed that a blackbody plus
power law model often provides a better fit than a single power
law. The blackbody temperature was always of $\sim$0.4-0.5 keV and
the power law photon index $\Gamma\sim$2 (except for the only
BeppoSAX observation carried out before the 1999 giant flare, that
had a harder spectrum with $\Gamma$=1.1). The XMM-Newton best fit
parameters are in agreement with these values, but the flux of
$\sim$$4.8\times10^{-12}$ \flux ~measured in our September 2005
observation is the lowest ever detected from \src. A $\sim$30\%
decrease of the persistent emission, compared to the
``historical'' level of $\sim$$10^{-11}$ \flux, had already been
noticed in the last BeppoSAX observation \citep{esposito06}, that
was carried out in April 2002, six months earlier than the last
bursts reported before the recent reactivation. This is
illustrated in Fig.~\ref{multi}, where we have plotted the long
term evolution of the pulse period, X--ray flux and bursting rate
of \src.

The long term fading experienced by \src\ in 2002-2005 might be
related to the apparent decrease in the bursting activity in this
period and can be compared to that of SGR 1627--41. SGR~1627--41
experienced a short period of bursting activity in June-July 1998
and, during the following $\sim$2 years, its 2-10 keV flux
decreased with time as a power law
F(t)$\propto$(t$-$t$_0$)$^{-0.6}$, with t$_0$ indicating the time
of the outburst \citep{mereghetti06}. As suggested by
\citet{kouveliotou03}, this behavior is likely due to the fact
that, during outbursts, a substantial amount of energy is
deposited  in the deep crustal layers ($\sim500-600$ m in depth)
due to shear dissipation and magnetic reconnection. Heat is then
transported inwards (because the conductivity increases at larger
densities) and later gradually transferred to the surface.
\citet{lyubarsky02} computed the surface cooling evolution, in
plane parallel approximation, by assuming a constant magnetic
field perpendicular to the surface and by solving numerically the
heat flow equation. They found that in a time scale of a few days
the deep crustal layers are cooled by inward heat flow, and that
80\% of the deposited energy is transferred to the core and
re-radiated over longer timescales as surface X-ray emission.
Quite independently on the details of the initial energy
deposition, this model gives a cooling luminosity that scales in
time as $t^{-0.7}$, in agreement to what has been observed for SGR
1627--41 \citep{mereghetti06}.

An alternative scenario to explain the``afterglows'' following
magnetars outbursts is that surface heating is caused by the
currents flowing in an azimuthally twisted magnetosphere
\citep{tlk02,gotthelf05}. The basic idea is that the toroidal
component of the internal magnetic field stresses the crust,
inducing a deformation and causing the external field to acquire
an azimuthal component. In this case a current density in excess
of the Goldreich-Julian current (which is expected for a simple
dipolar field) is required to thread the magnetosphere. As the
twist angle grows, the bursting activity is expected to increase
and larger returning currents heat the star surface producing more
thermal photons. By assuming a simple cylindrically symmetric and
self-similar magnetosphere, \citet{tlk02} derived an upper limit
for the luminosity of the returning currents, $L_X^{rc}\simeq
10^{35} B^p_{14} \Delta \phi$~erg s$^{-1}$, where $\Delta \phi$ is
the twist angle and $B^p_{14}$ is the polar value of the magnetic
field in unit of $10^{14}$~G.  In this model the luminosity decay
is dictated by the time evolution of the current (and that of the
consequent, almost instantaneous surface heating), but no detailed
computations have been performed so far.

The luminosity decay shown by \src\ has been much smaller than
that of SGR 1627--41, since the flux of the former  source only
faded by a factor $\sim$2 in three years. By fitting the observed
decay with a power law gives $F(t) \propto (t-t_0)^{-0.17}$, where
we have taken as t$_0$ the time of the intermediate flare of 18
April 2001  \citep{feroci03}.  The flatter slope may be an
indication that a mechanism of the second kind (i.e. surface
heating by returning currents) is at work. On the other hand, the
one-dimensional model computed by \citet{lyubarsky02} assumed that
the internal magnetic field is essentially radial. There is now
increasing theoretical and observational evidence that strong
poloidal and toroidal components can be present in the neutron
star crustal magnetic field (see e.g. \citealt{geppert06} and
references therein). This affects dramatically the heat transfer,
that becomes strongly anisotropic. A strong magnetic field
channels the heat flow along its field lines and,  in the presence
of large meridional components, can produce large inhomogeneities
in the surface temperature distribution.  Moreover, toroidal
fields substantially limit the radial conductivity (heat
blanketing) forcing energy to be transferred into narrow regions
along the polar axis. Although no detailed computations are
available, we may argue that, by assuming that the initial energy
deposition per unit volume is the same,  crustal fields with large
poloidal and toroidal components might produce flatter power law
luminosity decays, due to a combination of a smaller emitting
surface area and of the lower efficiency of the radial
conductivity in establishing a substantial thermal gradient
between the core and the surface (the latter being proportional to
the flux of heat outward).

We found no evidence for emission or absorption lines in the X-ray
spectra. The upper limits obtained in the longer observation of
September 2005 are the most constraining ever obtained for this
source in the ~1-10 keV energy range. An emission line at 6.4 keV
was possibly detected with the PCA instrument on RXTE in August
1998 \citep{strohmayer00}. This line,  visible only for the first
0.3 s of a particularly long and hard  burst, had an equivalent
width of $\sim$400 eV and was  interpreted as Fe fluorescence from
relatively cool material possibly ejected during the giant flare
that occurred two days before its detection. Thus it is not
surprising that we do not find evidence for the same feature in
the spectrum of the persistent emission.

In models involving ultra-magnetized neutron stars, proton
cyclotron features are expected to lie in the X-ray range, for
surface magnetic fields strengths of $\sim 10^{14}-10^{15}$~G.
Detailed calculations of the spectrum emerging from the
atmosphere of a magnetar in quiescence have confirmed this basic
expectation \citep{zane01,ho01}. Model spectra exhibit a strong
absorption line at the proton cyclotron resonance, $E_{c,p}\simeq
0.63 z_G(B/10^{14}\, {\rm G})$ keV, where $z_G$, typically in the
0.70--0.85 range, is the gravitational red-shift at the neutron
star surface. However, no evidence for persistent cyclotron
features have been reported to date in SGRs, despite some features
have been possibly detected during bursts (see e.g.
\citealt{strohmayer00,ibrahim03}).

Indeed some reasons have also been proposed to explain the absence
of cyclotron lines in magnetars, besides the obvious possibility
that they lie outside the sampled energy range. First, it must be
noticed that the atmospheric models available so far only account
for a single temperature and a single value of magnetic field
strength and inclination in the atmosphere, and no source of
heating besides the standard core cooling is taken into account.
Again, the lack of a standard atmosphere in active magnetars, and
the fact that their magnetic field topology and surface
temperature distribution are likely to be complex, makes the non
detection of proton cyclotron features in the persistent emission
not surprising.

Moreover, in the model discussed by \citet{tlk02}, magnetars  have
highly twisted magnetospheres that can support current flows.
These, in turn, can substantially distort the thermal emission
from the neutron star surface. The presence of charged particles
($e^-$ and ions) produces a large resonant scattering depth and
the resonant frequency depends on the local value of the magnetic
field. If the source flux at the cyclotron resonance does not
exceed the luminosity of the returning currents, the distributions
of both electrons and ions are spatially extended, in which case
repeated scatterings could lead to the formation of a hard tail,
typically observed below $\sim 10$~keV, instead of a narrow line.
Another implication of this model is that the twisted
magnetospheres can act as a source of gamma rays, either through
bremsstrahlung from a thin turbulent layer of the star's surface
heated to kT$\sim$100 keV by magnetospheric currents or through
synchrotron emission from pairs produced at a height of $\sim100$
km above the neutron star \citep{thompson05}. Indeed a hard
X--ray tail extending to 100 keV has been recently discovered in
\src\ with the INTEGRAL satellite \cite{gotz06}. A different
explanation for the absence of lines involves vacuum polarization
effects. It has been calculated that in strongly magnetized
atmospheres this effect can significantly reduce the equivalent
width of cyclotron lines, thus making difficult their detection
\citep{ho03}.

\section{Conclusions}

Thanks to the high sensitivity of the EPIC instrument on
XMM-Newton we have obtained the first high quality spectra of the
persistent X--ray emission from \src , setting tight limits on the
presence of emission and absorption lines.  In September 2005 the
source was found at a luminosity level of 1.3$\times10^{35}$ \lum\
(for d=15 kpc), a factor two smaller than the typical value
observed in the past, and in line with the trend of  luminosity
decrease already observed in the latest BeppoSAX observations
performed in April 2002. The target of opportunity XMM-Newton
observation of April 2006 showed that the decreasing luminosity
trend in \src\ has been interrupted by the recent onset of bursts
emission.
However, the moderate flux increase was not associated with
significant changes in the X--ray spectral and timing properties,
probably because the source is, up to now, only moderately active.
Future  observations with XMM-Newton will be essential to monitor
the spectral and flux variations for this source, possibly in
connection with its renewed bursting activity, as it has been
successfully done for its twin source SGR 1806--20.

We thank N.Schartel and the staff of the XMM-Newton Science Operation
Center for performing the target of opportunity observation.
This work has been partially supported by the Italian Space Agency
and INAF through contract ASI/INAF I/023/05/0
and by the MIUR under grant PRIN 2004-023189.

\bibliographystyle{aa}
\bibliography{biblio}

\clearpage

\begin{deluxetable}{cccccccccc}
\rotate
 \tabletypesize{\scriptsize}
  \tablecolumns{2}
\tablewidth{0pc}
 \tablecaption{Summary of the spectral results in
the 0.8--12 keV energy range\label{fits}}
\tablehead{\colhead{Model\tablenotemark{a}} &
\colhead{Observation} &  \colhead{$N_{\rm H}$} &
\colhead{$\Gamma$} & \colhead{$k_B T_1$} &
\colhead{$R_{bb\,1}$\tablenotemark{b}} &  \colhead{$k_B T_2$} &
\colhead{$R_{bb\,2}$\tablenotemark{b}} &  \colhead{Flux\tablenotemark{c}} &  \colhead{$\chi^{2}_{r}$ (d.o.f.)}\\
 \colhead{} & \colhead{} & \colhead{($10^{22}$ $\rm cm^{-2}$)} & \colhead{}  &
 \colhead{(keV)} & \colhead{(km)} & \colhead{(keV)} & \colhead{(km)}&
\colhead{($10^{-12}$ erg cm$^{-2}$ s$^{-1}$)} & \colhead{}}
\startdata
PL & A & $2.57\pm0.05$ & $2.84\pm0.04$ & \nodata & \nodata & \nodata & \nodata & $4.85\pm0.07$ & 1.65 (467) \\
\nodata & B & $2.71\pm0.08$ & $2.81\pm0.06$ & \nodata & \nodata & \nodata & \nodata & $5.6\pm0.1$ & 1.21 (323) \\
BB+BB  & A & $1.82\pm0.06$ & \nodata & $0.53^{+0.01}_{-0.02}$ & $3.7^{+0.3}_{-0.2}$ & $1.9\pm0.1$ & $0.22\pm0.02$ & $4.6\pm0.1$ & 1.32 (465) \\
\nodata & B & $2.0^{+0.1}_{-0.2}$ & \nodata & $0.53^{+0.03}_{-0.02}$ & $3.9\pm0.4$ & $1.9^{+0.2}_{-0.1}$ & $0.23^{+0.04}_{-0.03}$ & $5.3\pm0.2$ & 1.04 (321) \\
PL+BB & A &  $2.12\pm0.08$ & $1.9\pm0.1$ & $0.47\pm0.02$ & $4.0^{+0.4}_{-0.3}$ & \nodata & \nodata & $4.8\pm0.2$ & 1.24 (465) \\
\nodata & B &  $2.3^{+0.1}_{-0.2}$ & $1.9\pm0.2$ & $0.47\pm0.03$ & $4.2\pm0.5$ & \nodata & \nodata & $5.5\pm0.4$ & 1.00 (321) \\
\enddata
\tablenotetext{a}{Errors are quoted at the 90\% confidence level
for a single parameter.} \tablenotetext{b}{Radius at infinity
assuming a distance of 15 kpc.} \tablenotetext{c}{Flux in the
2--10 keV range, corrected for the absorption. The flux errors take into 
account the whole range of uncertainties in the spectral parameters.}
\end{deluxetable}

\clearpage

\clearpage

\clearpage

\clearpage

\clearpage

\clearpage

\end{document}